\documentstyle[epsfig]{mn}

\def\i{\item}
\def\bi{\bibitem{}}

\def\ni{\noindent}
\def\beb{\begin{thebibliography}{999}}
\def\eeb{\end{thebibliography}}
\def\bei{\begin{itemize}}
\def\eei{\end{itemize}}
\def\bef{\begin{figure}}
\def\eef{\end{figure}}
\def\ben{\begin{enumerate}}
\def\een{\end{enumerate}}
\def\beq{\begin{equation}}
\def\eeq{\end{equation}}
\def\ber{\begin{eqnarray}}
\def\eer{\end{eqnarray}}
\newcommand{\dmdt}{{\mbox{{\rm M}$_{\odot}$}} {\rm yr}$^{-1}$}

\newcommand{\mdot}{\mbox{$\dot{M}$}}

\newcommand{\lsim}{\raisebox{-0.3ex}{\mbox{$\stackrel{<}{_\sim} \,$}}}

\begin{document}
\title[neutron star field evolution]
{Magnetic Field Evolution of Accreting Neutron Stars - III}
\author[Konar and Bhattacharya]
{Sushan Konar$^{1,2,{\ast},{\dag}}$ and Dipankar Bhattacharya$^{1,{\dag}}$ \\
$^1$Raman Research Institute, Bangalore 560080, India \\
$^2$Joint Astronomy Program, Indian Institute of Science, Bangalore 560012, India \\
$^{\ast}$Present Address : IUCAA, Pune 411007, India \\
$^{\dag}$ e-mail : sushan@iucaa.ernet.in, dipankar@rri.ernet.in \\ }
\date{11th November, 1998}
\maketitle

\begin{abstract}
The evolutionary scenario of the neutron star magnetic field is examined assuming a spindown-induced 
expulsion of magnetic flux originally confined to the core, in which case the expelled flux undergoes 
ohmic decay. The nature of field evolution, for accreting neutron stars, is investigated incorporating
the crustal microphysics and material movement due to accretion. This scenario may explain the observed 
field strengths of neutron stars but only if the crustal lattice contains a large amount of impurity 
which is in direct contrast to the models that assume an original crustal field.
\end{abstract}


\section{Introduction}

\ni There is as yet no satisfactory theory for the generation of the neutron star magnetic field. There 
are two main possibilities - the field can either be a fossil remnant from the progenitor star, or be 
generated after the formation of the neutron star (for a review see Bhattacharya \& Srinivasan 1995, 
Srinivasan 1995 and references therein). Whereas post-formation generation mechanisms give rise to fields 
supported entirely by crustal currents (Blandford, Hernquist \& Applegate 1983), the fossil field resides 
in the core of a neutron star. In the core, the rotation of the star is supported by creation of Onsager-Feynman 
vortices in the neutron superfluid and the magnetic flux is sustained by Abrikosov fluxoids in the proton 
superconductor (Baym, Pethick \& Pines 1969, Ruderman 1972, Bhattacharya \& Srinivasan 1995). \\

\ni Evidently no consensus regarding the theory of field evolution has been reached since this depends crucially 
upon the nature of the underlying current configuration. Observations, though, indicate that the magnetic field 
decays significantly only if the neutron star is in an interacting binary (Bailes 1989, Bhattacharya 1991, Taam 
\& van den Heuvel 1986). There have been three major theoretical endeavours to link the field evolution with the 
binary history of the star - a) expulsion of the magnetic flux from the superconducting core during the phase of 
propeller spin-down, b) screening of the magnetic field by accreted matter and c) rapid ohmic decay of crustal 
magnetic field as a result of heating during accretion (see Bhattacharya 1995, 1996, Bhattacharya \& Srinivasan 
1995, Ruderman 1995 for detailed reviews). \\

\ni Except screening the other two models of field evolution depend on ohmic decay of the underlying current 
loops for a permanent decrease in the field strength. Such ohmic dissipation is possible only if the current 
loops are situated in the crust where the electrical conductivity is finite.  Models that assume an initial 
core-field configuration, therefore, require a phase of flux expulsion from the core.  Muslimov \& Tsygan 
(1985) and Sauls (1989) showed that there is likely to be a strong inter-pinning between the proton fluxoids and 
the neutron vortices. In a spinning down neutron star the neutron vortices migrate outward and by virtue of the 
inter-pinning drag the proton fluxoids along to the outer crust. Srinivasan et al.~(1990) pointed out that neutron 
stars interacting with the companion's wind would experience a major spin-down, causing the superconducting core 
to expel a large fraction of the magnetic flux. The nature of such flux expulsion as a result of spin-evolution 
has been investigated in detail for both isolated pulsars (undergoing pure dipole spin-down) and for the neutron 
stars that are members of binaries (Ding, Cheng \& Chau 1993; Jahan Miri \& Bhattacharya 1994; Jahan Miri 1996).
Recently, Ruderman, Zhu \& Chen (1998) has investigated the outward (inward) motion of core superfluid neutron 
vortices during spin-down (spin-up) of a neutron star which might alter the core's magnetic field in detail. \\

\ni Jahan Miri \& Bhattacharya (1994) and Jahan Miri (1996) have investigated the link between the magnetic field
evolution and the rotational history of a neutron star due to the interaction of its magnetosphere with the stellar 
wind of its companion. They assumed an uniform ohmic decay time-scale in the crust irrespective of the accretion rate.
Later, Bhattacharya \& Datta (1996) incorporated the crustal micro-physics into their calculation of the ohmic 
diffusion of an expelled field. However this work did not include the material movement that takes place in the crust 
as a result of accretion. Assuming that the field evolution starts only after the process of flux expulsion is over,
in the present work we incorporate the material movement and investigate the evolution of an expelled field in the 
crust of an accreting neutron star.

\section{Results and Discussions}

\ni Using the methodology developed by Konar \& Bhattacharya (1997 - paper-I hereafter) we solve the induction equation for 
an initial flux just expelled at the core-crust boundary due to spin-down. As in previous studies we solve this equation by 
introducing the vector potential $\vec A = (0, 0, A_\phi)$, where $A_\phi = g(r,t) \sin \theta / r$; $(r,\theta,\phi)$ being 
the spherical polar co-ordinates, assuming the field to be purely poloidal. For our calculation we have assumed an initial 
$g$-profile following the profile used by Bhattacharya \& Datta (1996) plotted in figure [\ref{fcore_gprofile}]. The sharply peaked nature of the $g$-profile is
caused by the flux expelled from the entire core being deposited in a thin layer,
thereby substantially increasing the local field strength.  The evolution 
of the crustal currents depend on the following parameters :
\bei
\i depth at which the currents are concentrated,
\i width of the current distribution,
\i impurity content of the crust,
\i rate of accretion.
\eei
In the present work we assume the flux to be deposited at the bottom of the crust and therefore the depth of the initial 
current configurations is taken to be the thickness of the crust. The evolution of such flux is not very sensitive to the
width of the current distribution (Bhattacharya \& Datta 1996). Therefore we keep the width of current distribution fixed
for all of our calculations. For details of the computation, crustal physics and binary parameters see Konar \& 
Bhattacharya (1999 - paper-II hereafter).  As in earlier papers, we denote
the impurity strength by the parameter 
\beq
Q \equiv \sum_i \frac{n_i}{n} (Z_i-Z)^2 \nonumber
\eeq
where $n$ and $Z$ are the number density and charge of background ions in the
pure lattice and $n_i$ and $Z_i$ those of the $i$-th impurity species.  The sum
extends over all species of impurities.
\bef
\begin{center}{\mbox{\epsfig{file=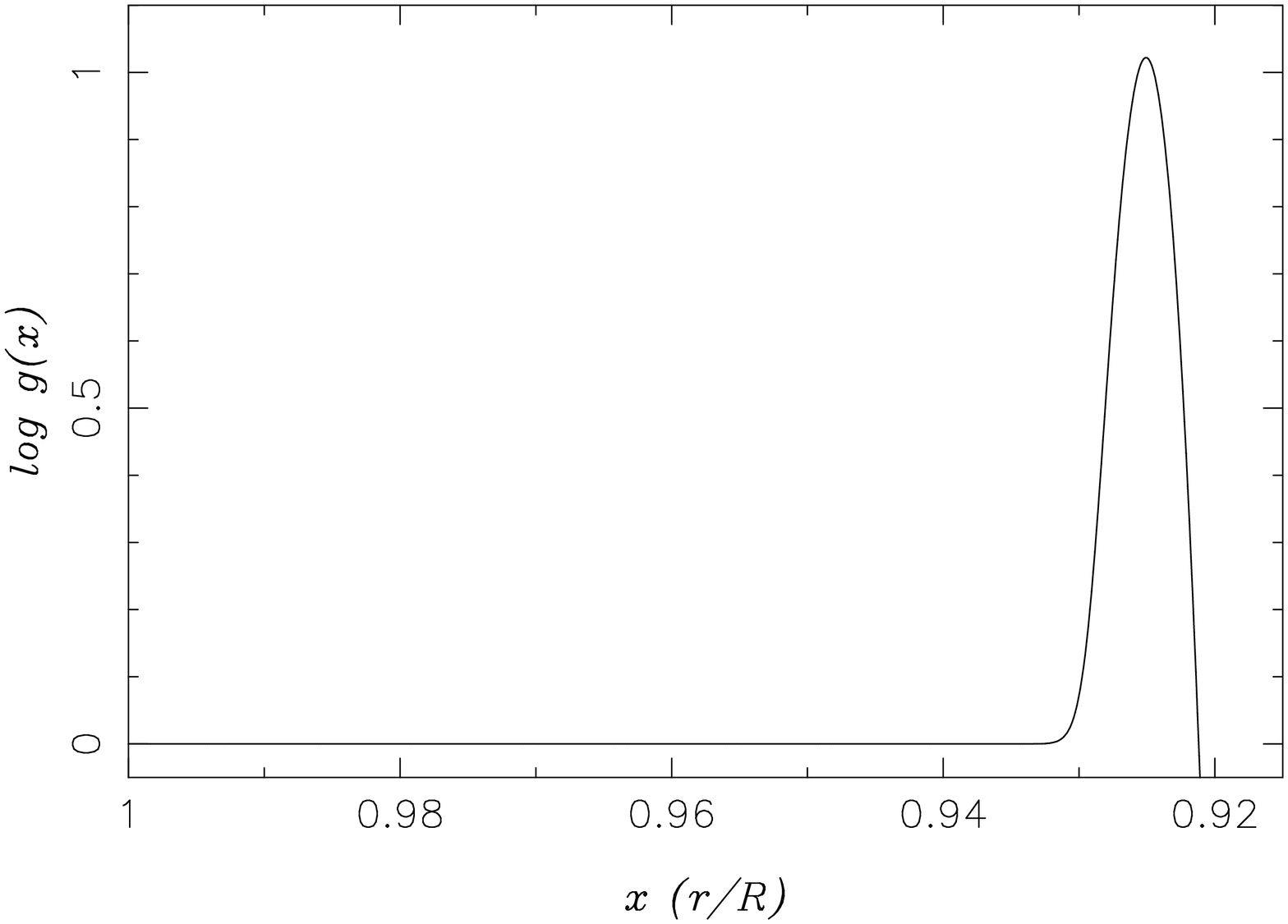,width=175pt,angle=-90}}}\end{center}
\caption[]{Initial radial $g$-profile in the crust immediately after expulsion.}
\label{fcore_gprofile}
\eef

\subsection{Field Evolution with Uniform Accretion}

\bef
\begin{center}{\mbox{\epsfig{file=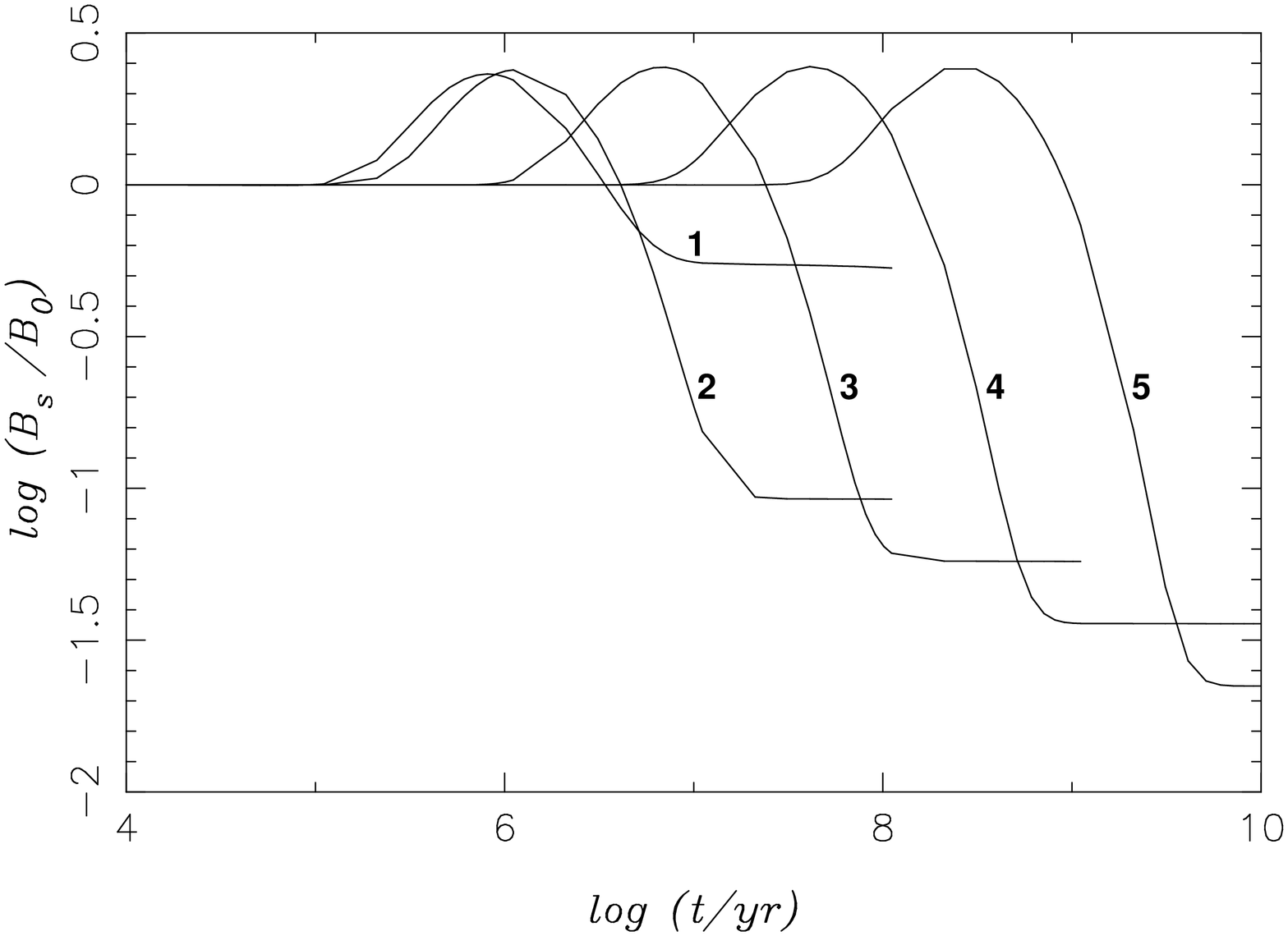,width=175pt,angle=-90}}}\end{center}
\caption[]{Evolution of the surface magnetic field for an expelled flux. The curves 1 to 5 correspond 
to $\mdot = 10^{-9}, 10^{-10}, 10^{-11}, 10^{-12}, 10^{-13}$~\dmdt. All curves correspond to $Q$ = 0.0.}
\label{fcore_accretion_dm}
\eef

\ni In figure [\ref{fcore_accretion_dm}] we plot the evolution of the surface field for different values of the 
accretion rate. The field strengths go down by about only an order and a half in magnitude even for fairly large 
values of the impurity strength. The characteristic features of field evolution, with uniform accretion, are as follows.
\ben
\i An initial rapid decay (ignoring the early increase) is followed by a slow down and an eventual {\em freezing}.
\i The onset of `freezing' is faster with higher rates of accretion.
\i Lower final `frozen' fields are achieved for lower rates of accretion.
\i To achieve a significant reduction in the field strength, very large values of the impurity strength are required.
\een
Hence, the general nature of field evolution in the case of an expelled flux is qualitatively similar to that in the 
case of an initial crustal flux (paper-I). However, to begin with, the expelled flux is deposited at the bottom of 
the crust. Therefore the initial rapid decay that is observed in case of crustal field is not as dramatic. Moreover, 
for higher rates of accretion the `freezing' happens much earlier. For example, with an accretion rate of $10^{-9}$~\dmdt 
the original crust is entirely assimilated into the core in about $10^7$ years. But the surface field levels off long 
before that since in this case the currents start returning to high density regions even before they have had time to 
spread out in the outermost regions of the crust. Due to the same reason large values of $Q$ do not change the final 
surface field much for higher rates of accretion, as for high enough temperatures and high densities the conductivity 
ceases to be sensitive to the impurity content of the crust (see paper-I for details). \\

\subsection{Field Evolution in Binaries}

\ni In paper-II we have considered three phases of binary evolution, namely - the isolated, the wind and the Roche-contact 
phase. In the wind phase there are two distinct possibilities of interaction between the neutron star and its companion. If 
the system is in the `propeller phase' then there is no mass accretion. But this phase is important because the star rapidly 
slows down to very long periods and as a result a significant flux-expulsion is achieved. From the point of view of flux 
expulsion, therefore, we assume the flux to be completely contained within the superconducting core (neglecting the small 
flux-expulsion caused by the dipole spin-down in the isolated phase) prior to this phase. The ohmic decay is assumed to take 
place only after this phase is over - that is in the phase of wind-accretion and in the phase of Roche-contact. In case of 
low mass X-ray binaries, it is not very clear as to how long the phase of wind-accretion lasts or whether such a phase is at 
all realized after the `propeller phase' is over. Therefore, in our calculations we have considered cases with and without a 
phase of wind accretion. \\

\subsubsection{High Mass Binaries}

\bef
\begin{center}{\mbox{\epsfig{file=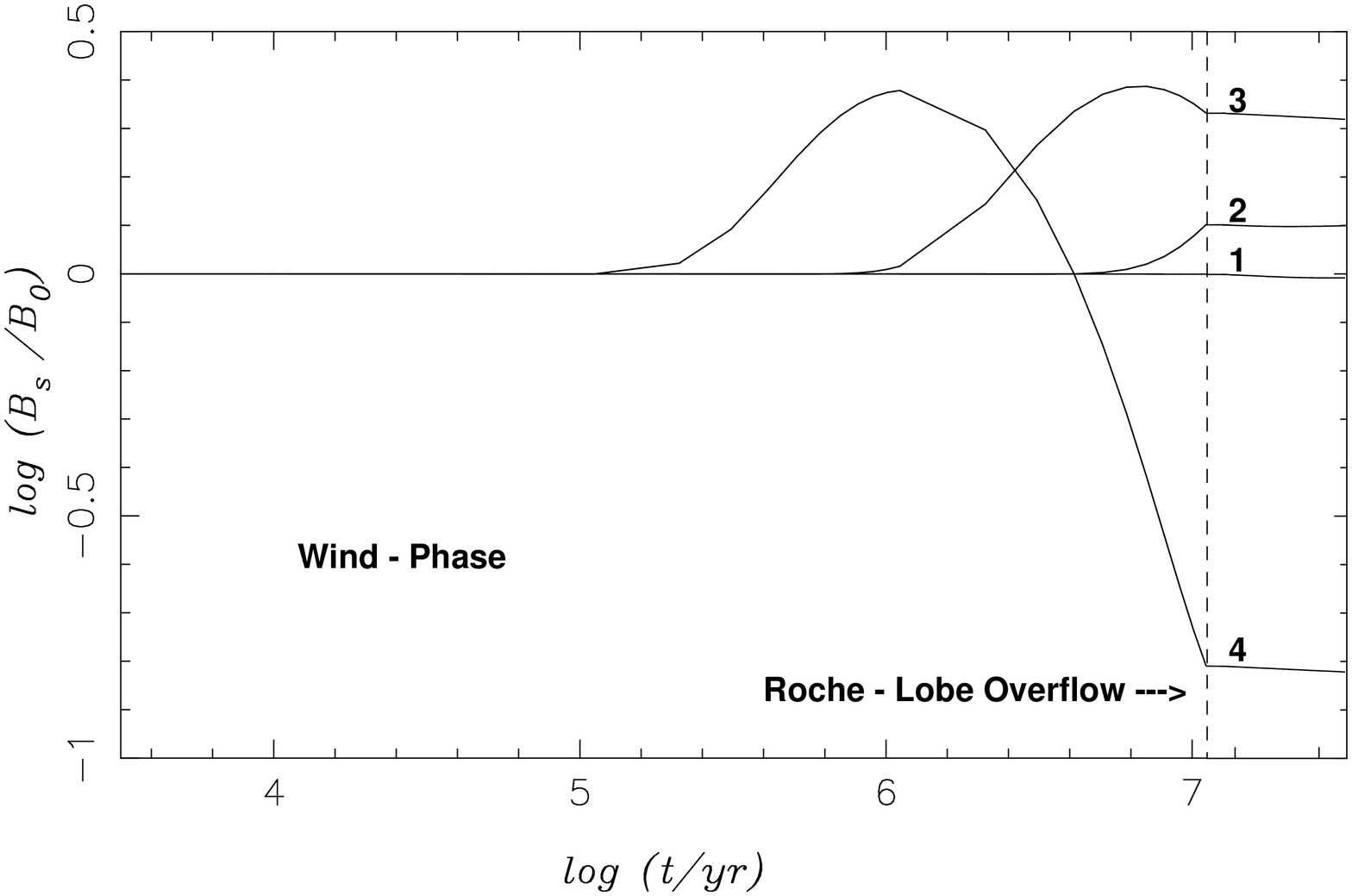,width=175pt,angle=-90}}}\end{center}
\caption[]{Evolution of the surface magnetic field in high mass X-ray binaries for four values of wind accretion rate. 
The curves 1 to 4 correspond to $\mdot = 10^{-13}, 10^{-12}, 10^{-11}, 10^{-10}$~\dmdt. All curves correspond to $Q$ = 0.0.}
\label{fcore_hmxb}
\eef

\ni Figure [\ref{fcore_hmxb}] shows the evolution of the surface field in high mass X-ray binaries for different rates of 
accretion in the wind phase. The surface field shows an initial increase. This is followed by a sharp decay (of about an
order of magnitude) only for sufficiently large rates of accretion in the wind phase. The decay in the Roche-contact phase 
is very small. In fact, the decay as seen in figure [\ref{fcore_hmxb}] is interrupted by Roche-contact in which the 
currents are quickly pushed to the core thereby `freezing' the field. This process also results in lower final 
field values for higher rates of accretion in the wind phase. Figure [\ref{fcore_accretion_dm}] shows that the field indeed 
decays faster for higher rates of accretion to begin with but the `freezing' happens earlier for higher rates thereby making 
the final saturation field lower for lower rates of accretion. Due to the short-lived nature of the wind-phase in massive 
binaries the `freezing' takes place before this saturation field is attained, giving rise to a behaviour 
(lower final field strengths for higher rates of accretion) contrary to that 
seen in low-mass binaries. For accretion 
rates appropriate to high mass X-ray binaries the field decreases at most by an order of magnitude. This result is quite
insensitive to the impurity strength of the crust because the time available for the evolution before the currents are pushed 
back into the core is rather small. \\

\subsubsection{Low Mass Binaries}

\bef
\begin{center}{\mbox{\epsfig{file=plot_lmxb.ps,width=160pt,angle=-90}}}\end{center}
\caption[]{Evolution of the surface magnetic field in low mass X-ray binaries. The dotted and the solid curves correspond 
to accretion rates of $\mdot = 10^{-9}, 10^{-10}$~\dmdt in the Roche contact phase.  The curves 1 to 5 correspond to 
$Q$ = 0.0, 0.01, 0.02, 0.03 and 0.04 respectively. All curves correspond to a wind accretion rate of $\mdot = 10^{-16}$~\dmdt. }
\label{fcore_lmxb}
\eef

\bef
\begin{center}{\mbox{\epsfig{file=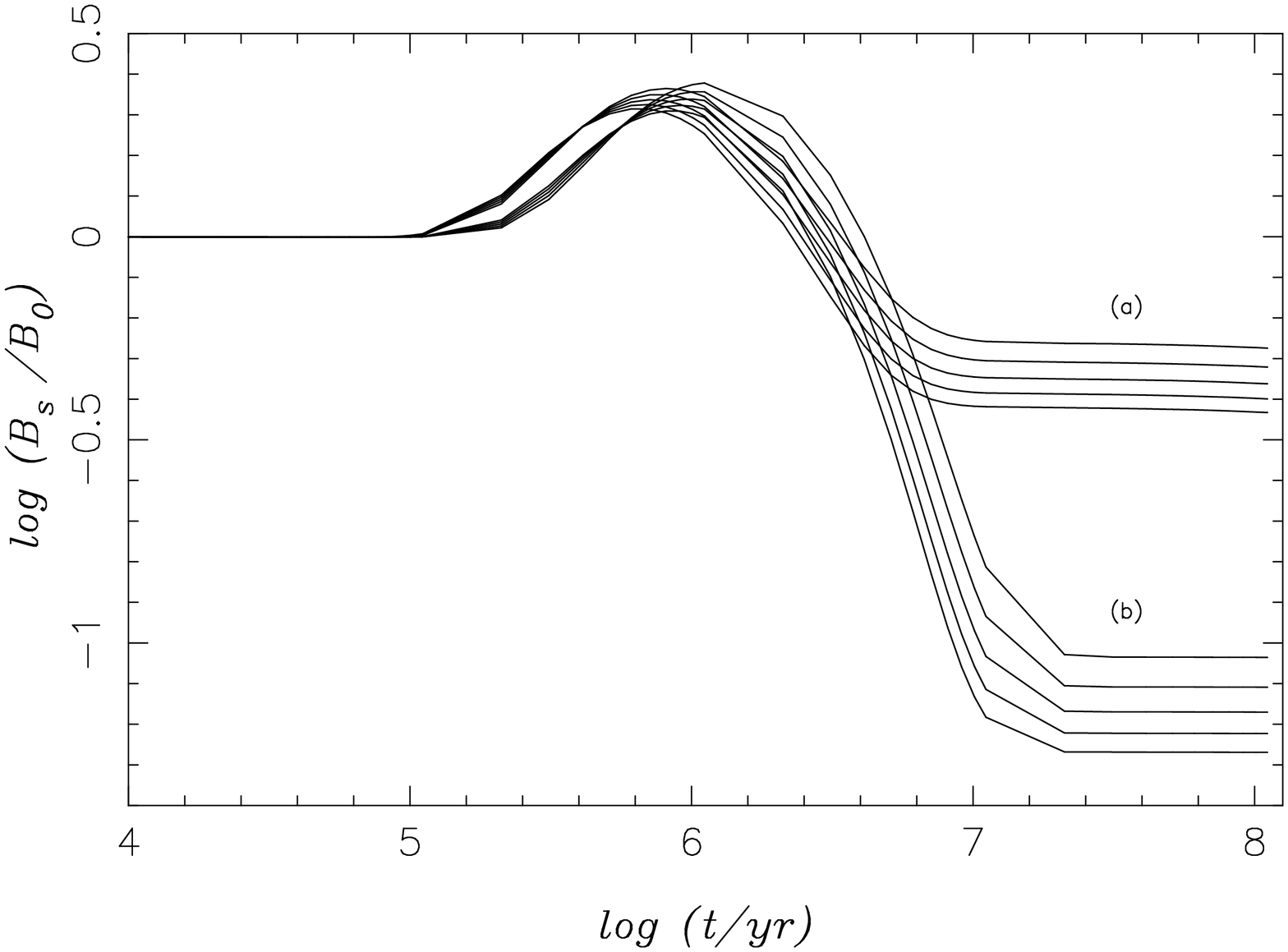,width=175pt,angle=-90}}}\end{center}
\caption[]{Evolution of the surface magnetic field in low mass X-ray binaries without a phase of wind accretion. 
The set of curves (a) and (b) correspond to accretion rates of $\mdot = 10^{-9}, 10^{-10}$~\dmdt in the Roche contact 
phase. Individual curves in each set correspond to $Q$ = 0.0, 0.1, 0.2, 0.3 and 0.4 respectively, the upper curves
being for the lower values of $Q$.} 
\label{fcore_lmxb_nw}
\eef

\ni Figure [\ref{fcore_lmxb}] shows the evolution of the surface field in low mass X-ray binaries, for different 
values of the impurity concentration in the crust. It should be noted that a difference in the wind accretion rate 
does not manifest itself in either the nature of the field evolution or the final field strength. However, a 
difference in the accretion rate in the Roche-contact phase shows up very clearly. Comparing the different curves 
(for different values of the impurity parameter) we see that a large value of impurity strength gives rise to a rapid 
initial decay and therefore a lower value of the final surface field.  \\

\ni In figure [\ref{fcore_lmxb_nw}] we have plotted the evolution of the surface field assuming the wind accretion
phase to be absent. Once again we find that for higher rates of accretion higher final field values are obtained. 
It should be noted here that the final field values obtained now are only about an order and a half of magnitude 
lower than the original surface field strengths. Even though the impurity strengths assumed now are much higher than 
those assumed for the field evolution in low mass X-ray binaries {\em with} a phase of wind-accretion. In absence of 
a prior phase of wind accretion the flux does not have enough time to diffuse out to low density regions when the 
Roche-contact is established. Therefore the role of accretion, in the Roche-contact phase, is mainly to push the 
currents towards the high density interior rather than to enhance ohmic decay rate. Evidently, much larger impurity 
strength is required for the final field values to decrease by three to four orders of magnitude. Unfortunately, 
our code is unable to handle high $Q$ at present because of prohibitive requirement of computer time to ensure
stability. However, this figure clearly establishes a trend as to how the final field values behave with $Q$ and it 
is evident that we need $Q$ values much larger than those considered here to achieve millisecond pulsar field strengths 
in systems {\em without} a phase of wind accretion. Alternatively, systems without wind accretion might lead to high
field pulsars in low-mass binaries, such as PSR 0820+02. \\

\ni The most important point to note here is that similar to an initial crustal field configuration, the amount of field 
decay is much larger than that achieved in the case of high mass X-ray binaries.  Although in low mass X-ray binaries the 
surface field does go down by three to four orders of magnitude from its original value for large values of impurity strength, 
the final field could remain fairly high if the impurity strength is small.  If the wind-accretion phase is absent in these 
systems then to achieve large amount of field reduction even higher values of the impurity strength becomes necessary. 
Therefore, the `spin-down induced flux expulsion model' will be consistent with the overall scenario of field evolution and 
in particular millisecond pulsars can be produced in low mass X-ray binaries provided the impurity strength in the crust of 
the neutron stars is assumed to be extremely large. \\

\ni The results of our investigation (the present work and that described in paper-II) clearly indicate that both the 
models - assuming an initial crustal field or, alternatively a spin-down induced flux expulsion, place very stringent 
limits on the impurity strength. Whereas for the crustal model only $Q \lsim 0.01$ is allowed, the results described 
here show that much larger values of impurity are needed in the latter model. These requirements are quite different. 
If there is an independent way of estimating the impurity content of the crust then we could differentiate between
these two models. However, in all of our investigation we have assumed that the impurity content of the crust does 
not change as a result of accretion, which may not be quite correct since accretion changes the crustal composition 
substantially (H\t{ae}nsel \& Zdunik 1990). \\

\section{conclusions}

\ni In this work we have investigated the consequences of `spin-down induced flux expulsion'. The general nature of field 
evolution seems to fit the overall scenario. The nature of field evolution is quite similar to that in the case of a purely 
crustal model of field evolution though the details differ. Most significantly, this model has the requirement of large 
values of the impurity strength $Q$ in direct contrast to the crustal model. To summarize then:
\bei
\i The field values in the high mass X-ray binaries can remain fairly large for a moderate range of impurity strength.
\i A reduction of three to four orders of magnitude in the field strength can be achieved in the low mass X-ray binaries 
provided the impurity strength is as large as 0.05.
\i If the wind accretion phase is absent then to achieve millisecond pulsar field values, impurity strength in excess 
of unity is required.
\eei

\beb
\bi Bailes M., 1989, ApJ, 342, 917 
\bi Baym G., Pethick C., Pines D., 1969, Nat, 223, 673
\bi Bhattacharya D., 1991, {\em Neutron Stars : Theory and Observations}, ed. Ventura E., Pines D., 
Kluwer Academic Publishers, p. 219
\bi Bhattacharya D., 1995, JA\&A, 16, 217
\bi Bhattacharya D., 1996, {\em Pulsar Timing, General relativity and 
the Internal Structure of Neutron Stars}, ed. Z. Arzoumanian et al, Royal Netherlands Academy of Arts and Sciences, in press.
\bi Bhattacharya D., Datta B., 1996, MNRAS, 282, 1059
\bi Bhattacharya D., Srinivasan G., 1995, {\em X-Ray Binaries}, ed. 
Lewin W.~H.~G., van Paradijs J., van den Heuvel E.~P.~J., Cambridge University Press, p. 495
\bi Blandford R.~D., Applegate J.~H., Hernquist L. 1983, MNRAS, 204, 1025
\bi Ding K.~Y., Cheng K.~S., Chau H.~F., 1993, ApJ, 408, 167
\bi H\t{ae}nsel P., Zdunik J.~L., 1990, A\&A, 229, 117
\bi Jahan Miri M., 1996, MNRAS, 283, 1214
\bi Jahan Miri M., Bhattacharya D., 1994, MNRAS, 269, 455 
\bi Konar S., Bhattacharya D., 1997, MNRAS, 284, 311
\bi Konar S., Bhattacharya D., 1999, MNRAS, 303, 588
\bi Muslimov A.~G., Tsygan A.~I., 1985, SvAL, 11, 80
\bi Ruderman M.~A., 1972, ARA\&A, 10, 427
\bi Ruderman M., 1995, JA\&A, 16, 207
\bi Ruderman M., Zhu T., Chen K., 1998, ApJ, 492, 267
\bi Sauls J., 1989, {\em Timing Neutron Stars}, ed. \'{O}gelman H., van den Heuvel E.~P.~J., 
Dordrecht, Kluwer, p. 457
\bi Srinivasan G., 1995, {\em Stellar Remnants}, Saas-Fee Advanced Course 25, ed. Meynet G., 
Sherev D., Springer-Verlag
\bi Srinivasan G., Bhattacharya D., Muslimov A.~G., Tsygan A.~I., 1990, Curr.Sc., 59, 31
\bi Taam R.~E., van den Heuvel E.~P.~J., 1986, ApJ, 305, 235
\eeb

\end{document}